\documentclass[openacc]{rsproca}

\usepackage{color}

\newcommand{\be}{\begin{equation}}
\newcommand{\ee}{\end{equation}}

\DeclareMathOperator{\sgn}{sgn} 

\newcommand{\mL}{\mathcal{L}}
\newcommand{\PP}{\mathbb{P}}
\newcommand{\RR}{\mathbb{R}}
\newcommand{\NN}{\mathbb{N}}
\newcommand{\CC}{\mathbb{C}}

\newcommand{\lLambda}{\widetilde{\Lambda}}
\newcommand{\llambda}{\widetilde{\lambda}}
\newcommand{\lfLambda}{\widehat{\widetilde{\Lambda}}}
\newcommand{\fk}{\widehat{k}}
\newcommand{\lPsi}{\widetilde{\Psi}}
\newcommand{\lpsi}{\widetilde{\psi}}
\newcommand{\lPsiM}{\widetilde{\Psi}_{\rm{M}}}
\newcommand{\lpsiM}{\widetilde{\psi}_{\rm{M}}}

\newcommand{\e}{{\rm e}}
\newcommand{\lk}{\left(}
\newcommand{\rk}{\right)}
\newcommand{\lkk}{\left\{ }
\newcommand{\rkk}{\right\} }

\begin{document}

\title{Sturm--Liouville systems for 
the survival probability in first-passage time problems}

\author{
M. Dahlenburg$^{1,2}$ and G. Pagnini$^{1,3}$}

\address{$^{1}$BCAM -- Basque Center for Applied Mathematics, 
Alameda de Mazarredo 14, 48009 Bilbao, Basque Country -- Spain\\
$^{2}$University of Potsdam, Institute for Physics \& Astronomy,
Karl-Liebknecht-St 24/25, 14476 Potsdam, Germany\\
$^{3}$Ikerbasque -- Basque Foundation for Science, 
Plaza Euskadi 5, 48009 Bilbao, Basque Country -- Spain
}
\Year{2023}
\Doi{2023.0485}

\citearticle{Dahlenburg M, Pagnini G}
\subject


\corres{Gianni Pagnini\\
\email{gpagnini@bcamath.org}}

\begin{abstract}
We derive a Sturm--Liouville system of equations for 
the exact calculation of the survival probability 
in first-passage time problems.
This system is the one associated with
the Wiener--Hopf integral equation obtained from the
theory of random walks.
The derived approach is an alternative to the existing literature
and we tested it against direct calculations 
from both discrete- and continuous-time random walks
in a manageable, but meaningful, example. 
Within this framework, the Sparre Andersen theorem 
results to be a boundary condition for the system.
\end{abstract}


\begin{fmtext}

\section{Introduction}
\label{sec:intro}
The probability for a random walker started at $a > 0$ to remain
in the initial half-axis after $n$ steps is
called survival probability \cite{bray_etal-ap-2013}. 
If we denote it by $\phi_n(a)$,
for $z , \xi \in \RR_0^+$, it holds
\cite{spitzer-1957,bray_etal-ap-2013}
\be
\int_0^\infty k(z-\xi) \, \phi_n(z) \, dz = \phi_{n+1}(\xi) \,,
\quad n \ge 0 \,,
\label{WH}
\ee
with initial condition $\phi_0(\xi)=1$, for all $\xi \in \RR_0^+$,
where $k(x)$, with $x \in \RR$, is the distribution of the 
jumps.
A general solution to \eqref{WH} is known in literature
with the name of Pollaczek--Spitzer formula \cite{bray_etal-ap-2013}.
This name refers to a formula, 
see \cite[theorem 5, formula (4.6)]{spitzer-1957} and
\cite[formula (12)]{bray_etal-ap-2013}, 
derived by F. Spitzer in 1957 \cite[theorem 3, formula (3.1)]{spitzer-1957}
on the basis of an auxiliary formula obtained by F. Pollaczek in 1952 
\cite[formula (8)]{pollaczek-1952} but through a different method. 
Here we derive the Sturm--Liouville system associated with
\eqref{WH} for the calculation  
\end{fmtext}

\maketitle

\noindent
of the survival probability $\phi_{n+1}(\xi)$
as an alternative and
easier approach with respect to the Pollaczek--Spitzer formula.
Thus, first we show that $\phi_{n+1}(\xi)$ is indeed
the solution of a differential equation
and later we check this solution against
direct calculations in a manageable, but meaningful, 
case within both the framework of discrete-time random walk
and also that of continuous-time random walk (CTRW).

When $a=0$, the survival probability $\phi_n(a)$ is indeed determined by
the celebrated Sparre Andersen theorem derived on the basis
of combinatorial arguments \cite{sparre-andersen-1954b}.
Within the provided setting, the Sparre Andersen theorem \eqref{SA} 
actually is a boundary condition of the Sturm--Liouville system
of equations in order to have a unique and non-decreasing solution 
$\phi_{n+1}(\xi)$. Sparre Andersen theorem is a fundamental result
in the study of first-passage time problems for 
Markovian symmetric random walks with discrete time-step and
starting at the origin.
Originally established in 1954 
\cite{sparre-andersen-1954b}, the theorem states
\be
\phi_n(0) 
= 2^{-2n} \binom{2n}{n}
\sim \frac{1}{\sqrt{n \pi}} \,, \quad n \to \infty \,,
\label{SA}
\ee
where $n \in \NN_0$ is the epoch of a one-dimensional  
random walk $S_n=X_1 + \dots + X_n$ started at the origin $S_0=0$ 
with jumps i.i.d. random variables in $\RR$
and $\phi_n(0)$ is 
the probability for the first ladder epoch 
$\{\mathcal{T}=n\}=\{S_1 \ge 0, \dots, S_{n-1} \ge 0, S_n < 0\}$ 
to be larger than $n$, i.e., $\phi_n(0)=\PP(\mathcal{T} > n)$.
Thus, $\mathcal{T}$ is the epoch of the first entry of the walker
into the strictly negative half-axis.
See also F. Spitzer \cite{spitzer-1956} and 
W. Feller \cite[Section XII.7]{feller-1971}.
By adding a constant $a > 0$ to all terms, 
we obtain a random walk starting at $a$.  

In opposition to the present approach,
the derivation of the Sparre Andersen theorem
\eqref{SA} from \eqref{WH} in the limit $\xi \to 0$ is 
possible by taking into account the boundedness of 
the survival probability at infinity as a consequence
of its physical interpretation, see, e.g.,
\cite{frisch_etal-1975,frisch_etal-1994,
vergassola_etal-aa-1994}.
By the way, the authors did not 
calculate with that approach also the survival probability
\cite{frisch_etal-1975,frisch_etal-1994,vergassola_etal-aa-1994}. 
Sometimes, the discussion of the Pollaczek--Spitzer formula 
suggests that it is possible to get the survival probability 
from it together with the Sparre Andersen theorem,
see, e.g., \cite{majumdar-pa-2010,bray_etal-ap-2013},
but this is misleading.
Unfortunately, as reported by F. Spitzer in his proof
\cite[theorem 5]{spitzer-1957},
the so-called Pollaczek--Spitzer formula 
\cite[formula (4.6)]{spitzer-1957} requires 
two conditions. The first condition 
is provided by the following lemma \cite[lemma 8]{spitzer-1957}:

The non-decreasing limit function 
\be
\phi(\xi)=\lim_{n\to\infty}\frac{\phi_n(\xi)}{\phi_n(0)} \,, \quad
\xi > 0 \,,
\label{philimit}
\ee
satisfies the equation
\[
\phi(\xi)=\int_0^\infty k(z-\xi)\phi(z) \, dz \,, \quad
\xi > 0 \,,
\]
where point $\xi=0$ is excluded because it is assumed to be known 
\cite[formula (4.4)]{spitzer-1957}.
Later, through the same limit \eqref{philimit},
it requires, as second condition, 
also the estimation of the following limit \cite[formula (4.3)]{spitzer-1957}:
\[
\phi(\xi)=
\lim_{n \to \infty}\sqrt{n\pi} \, \phi_n(\xi) \,, 
\]
that uses formula \eqref{SA} as an external information
for quantifying $\phi_n(0)$, read - settled for this text -:
{\it "This case is rather simple, because it is then known 
(see \cite{sparre-andersen-1954b} or \cite{spitzer-1956}) that 
\eqref{SA}"} \cite[formula (4.4) and above lines]{spitzer-1957}. 
This means that claiming the derivation of the Sparre Andersen theorem 
\eqref{SA} from the Pollaczek--Spitzer formula is indeed misleading, 
see, for example, reference
\cite[from formula (12) to (13) without steps]{bray_etal-ap-2013}, 
because Pollaczek--Spitzer formula in the limit $\xi \to 0$
provides \eqref{SA} since it has been
already assumed for its derivation. 
However, a further derivation of the Pollaczek--Spitzer formula 
is available in literature 
that allows for obtaining the survival probability
together with the Sparre Andersen theorem 
by assuming the boundedness of the survival probability
\cite{ivanov-aa-1994,comtet_etal-jsm-2005,majumdar_etal-jsm-2014}.
This is analogue to the assumption on
the boundary conditions required in the approach here proposed.

The rest of the paper is organised as follows.
In section \ref{sec:sturmliouville}, 
we derive the Sturm--Liouville system associated 
with the Wiener--Hopf equation \eqref{WH}.
In section \ref{sec:DTRW}, we discuss our result through
a meaningful example from the theory of the discrete-time random walk
and in section \ref{sec:CTRW} through the same example 
from the theory of the CTRW. 
Finally, in section \ref{sec:conclusions} the 
paper is summarised and conclusions are reported.

\section{The associated Sturm--Liouville system}
\label{sec:sturmliouville}
We are now interested in solving \eqref{WH} with a method that is
different from the Pollaczek--Spitzer formula.
First we observe that $z,\xi \in \RR_0^+$ while $x=(z-\xi) \in \RR$
and $\displaystyle{\int_{\RR}k(x)dx=1}$.
Moreover, a noteworthy property of \eqref{SA} is that it is independent
of the distribution of the jumps if a symmetric kernel $k(x)=k(|x|)$
is considered. 
Therefore, in spite of the fact that the kernel 
$k(z,\xi)=k(z-\xi)=k(|z-\xi|)$ 
is assumed to be everywhere continuous in $z$,
its derivative with respect to $z$ 
has a jump discontinuity at $z=\xi$, i.e.,
\be
k'_z(z,\xi)=\frac{dk(|z-\xi|)}{dz}
= \sgn(z-\xi) \,
\frac{dk(|x|)}{d|x|} \,,
\label{kdev}
\ee
and we define
\be
-\frac{1}{p(\xi)} = k'_z(\xi^+,\xi)-k'_z(\xi^-,\xi) \,.
\label{discontinuity}
\ee

There are some general classes of boundary value problems
for differential equations that can be reduced to integral equations
\cite{kisil_etal-prsa-2021}.
More specifically, here we consider the reduction of the
Sturm--Liouville systems to 
Fredholm integral equations of the second kind. 
Thus, by following F.G. Tricomi \cite[Chapter 3.13]{tricomi-1985},
we introduce an operator $\mL$ such that
\be
\mL[k(z,\xi)] 
= \frac{d}{dz}\left[
p(z)\frac{dk}{dz}\right] + q(z)k(z,\xi) 
= 0 \,, \quad z\ne\xi\,,
\label{Lk}
\ee
and 
\be
\mL[\phi_{n+1}(z)]
= \frac{d}{dz}\left[
p(z)\frac{d\phi_{n+1}}{dz}\right] + q(z)\phi_{n+1}(z)
=-\phi_n(z) \,, 
\label{Lphi}
\ee
for an appropriate function $q(z)$
and for a source term $\phi_n(z)$,
together with the corresponding boundary conditions 
at $z=0$ and $z=\infty$ of $k(z,\xi)$, $\phi_{n+1}(z)$
and their first derivative.
Since the distribution $k(x)$ of the jumps 
is assumed to be known, 
function $q(z)$ is given by the choice of $k(x)$,
and $\phi_n(z)$ is indeed iteratively determined from 
the initial condition $\phi_0(z)=1$.
Then, from the subtraction 
$k(z,\xi)\mL[\phi_{n+1}(z)] - \phi_{n+1}(z)\mL[k(z,\xi)]
=-k(z,\xi)\phi_n(z)$ 
we remove $q(z)$ and we have the Sturm--Liouville system
\be
\frac{d}{dz}\left[
p(z)\frac{d\phi_{n+1}}{dz}\right]=
\frac{\phi_{n+1}(z)}{k(z,\xi)}
\frac{d}{dz}\left[
p(z)\frac{dk}{dz}\right] 
- \phi_n(z) \,.
\label{SL}
\ee
We now show that \eqref{SL} is 
the Sturm--Liouville system associated with the Wiener--Hopf integral equation
\eqref{WH} for proper boundary conditions.

We start by considering the integral
\begin{eqnarray}
I &=& \int_0^\infty k(z,\xi)\mL[\phi_{n+1}(z)] \, dz \nonumber \\
&=& I_0 + \int_0^\infty k(z,\xi) q(z)\phi_{n+1}(z) \, dz \,,
\label{I}
\end{eqnarray}
where the second line follows from the LHS of \eqref{Lphi} with
\be
I_0=
\int_0^\infty k(z,\xi) \frac{d}{dz}\left[
p(z)\frac{d\phi_{n+1}}{dz}\right] \, dz 
\,.
\label{I0}
\ee
From the RHS of \eqref{Lphi} we also have that
\be
I = - \int_0^\infty k(z,\xi) \phi_{n}(z) \, dz \,.
\label{IRHS}
\ee
Integrating by parts $I_0$ \eqref{I0}, it holds
\be
I_0 =
\left. k \, p \, \frac{d\phi_{n+1}}{dz}\right|_0^\infty
- \int_0^\infty k'_z \, p \, \frac{d\phi_{n+1}}{dz} \, dz \,,
\label{I02}
\ee
thus, by integrating again by parts the integral in the
RHS of \eqref{I02} and reminding \eqref{Lk}, 
integral \eqref{I} turns into
\be
I =
\left. k \, p \, \frac{d\phi_{n+1}}{dz}\right|_0^\infty
- \int_0^\infty \frac{d}{dz}\left[k'_z \, p \, \phi_{n+1}\right] \, dz 
\,.
\label{I2}
\ee
Because of jump discontinuity of $k'_z$ \eqref{discontinuity},
the integral term on the RHS of \eqref{I2} gives
\be
\int_0^\infty = \int_0^{\xi^-} + \int_{\xi^+}^\infty =
\left. \frac{dk}{dz} \, p \, \phi_{n+1} \right|_0^\infty
+ \phi_{n+1}(\xi) \,,
\ee
and, finally, by comparing \eqref{I2} and \eqref{IRHS} we recover
\eqref{WH} provided that 
\be
\left[p(z)\left(k(z,\xi)\frac{d\phi_{n+1}}{dz}
- \phi_{n+1}(z)\frac{dk}{dz}\right)\right]_0^\infty
= 0
\,.
\label{BC}
\ee
Hence, the kernel $k(z,\xi)$ and the survival probability $\phi_{n+1}(z)$ 
which are related by the Wiener--Hopf integral equation \eqref{WH} are indeed 
the solutions of equations \eqref{Lk} and \eqref{Lphi}, or \eqref{SL},
when boundary conditions \eqref{BC} are met.
We highlight that in \eqref{BC}, the kernel $k(z,\xi)$ is supposed to be
known from the process and $p(z)$ is determined through \eqref{discontinuity}.
Thus, boundary conditions \eqref{BC} define indeed the boundary values
of $\phi_{n+1}(z)$ and its first derivative. 
In the general case when boundary conditions \eqref{BC} are different
from $0$, this procedure reduces differential equations
to Fredholm integral equations of the second kind. 
In our case \eqref{BC}, 
this procedure reduces differential equation \eqref{SL}
to a homogeneous Fredhdolm integral equation of second kind,
that is the Wiener--Hopf integral equation \eqref{WH}.

\section{An example from the theory of discrete-time random walks}
\label{sec:DTRW}
Wiener--Hopf integral equation \eqref{WH} 
provides the survival probability in the case of a discrete-time
random walk. In this section,
first we directly solve \eqref{WH} for the generating function
by means of classical methods in complex analysis for the 
meaningful and manageable example with 
the exponential kernel $k(x)=\e^{-|x|}/2$, 
see, e.g., \cite{gutkowiczkrusin_etal-jsp-1978,kou_etal-aap-2003,
majumdar_etal-jsp-2006,dahlenburg_etal-jpa-2022}, and later 
we solve the Sturm--Liouville system derived in section \ref{sec:sturmliouville}
for the same special case.

We introduce the generating function
\be
\overline{\phi}(\xi,u) = \sum_{n=0}^\infty u^n \phi_n(\xi)
= 1 + \sum_{n=1}^\infty u^n \phi_n(\xi) \,, 
\label{genfun}
\ee
which solves the equation
\be
\overline{\phi}(\xi,u) = 1 + u\int_{0}^\infty 
k(z-\xi) \overline{\phi}(z,u) \, dz \,.
\label{WHgenfun}
\ee
Equation \eqref{WHgenfun} is a Fredholm integral equation
of the second kind with respect to $z$ and it can be solved
by methods developed for Wiener--Hopf integral equations 
\cite{estrada_kanwal-2000}.
Thus, by following Wiener--Hopf technique, 
we extend the interval of $\xi$ to $\RR$ and we
refer to $\overline{\phi}_+(\xi,u)$ and $\overline{\phi}_-(\xi,u)$ 
for {\it positive} and {\it negative} values of $\xi$, respectively,
such that $\overline{\phi}=\overline{\phi}_+ + \overline{\phi}_-$.
We are interested to find $\overline{\phi}_+(\xi,u)$ that corresponds
to the physical solution of \eqref{WHgenfun} while $\overline{\phi}_-(\xi,u)$ 
is an auxiliary function. 
We define the generalised Fourier transform with $\omega \in \CC$
and we have the pairs
\be
\widehat{g}_\pm(\omega) = \pm \int_0^{\pm\infty} \e^{i\omega x}g_\pm(x) \, dx \,,
\label{fourier}
\ee
\be
g_\pm(x) = \frac{1}{2\pi} \int_{L_\pm} \e^{- i\omega x}
\widehat{g}_\pm(\omega) \, d\omega \,,
\label{fourierinv}
\ee
where $L_\pm$ are proper integration paths in the complex plane.
By inverting the Fourier transform,
we explicitly obtain the generating function as follows
\cite{kisil_etal-prsa-2021}
\begin{eqnarray}
\overline{\phi}_{+}(\xi,u)
&=& \frac{1}{2\pi}\int_{L_{+}} 
\e^{-i\omega \xi} \, \widehat{\overline{\phi}}_{+}(\omega,u) \, d\omega \,,
\nonumber \\
&=&  - {\rm Res} \lkk \frac{C \, \overline{\phi}_{-}(0,u)(i+\omega)
\, \e^{-i\omega \xi}}{1+\omega^2-u} - 
\frac{(1+\omega^2) \, \e^{-i\omega \xi}}{\omega\lk 1+\omega^2-u\rk} \rkk \,,
\nonumber \\
&=& \frac{-C \overline{\phi}_{-}(0,u)}{2} \lkk 
\frac{(\sqrt{1-u}-1) \exp(-\sqrt{1-u} \, \xi)}{\sqrt{1-u}}
+ \frac{(1+\sqrt{1-u})\exp(\sqrt{1-u}\, \xi)}{\sqrt{1-u}} \rkk + \nonumber\\
& & \quad + \frac{1}{1-u}+\frac{u \,\exp(-\sqrt{1-u} \,\xi)}{2(u-1)} 
+ \frac{u \, \exp(\sqrt{1-u} \, \xi)}{2(u-1)} \,.
\label{lambda_d_not_uni}
\end{eqnarray} 
where $C$ is a multiplicative constant that appears 
in the solution of the homogeneous case.
Since the generating function is bounded because of its
physical meaning, i.e.,
\begin{equation}
\label{bounded}
\lim_{\xi \to \infty} \overline{\phi}_{+}(\xi,u)
= \overline\phi_\infty(u) < \infty \,,
\end{equation}
then from \eqref{lambda_d_not_uni} 
we have that the exponential term $\exp(\sqrt{1-u} \, \xi)$ must disappear,
which gives
\begin{equation}
C \, \overline{\phi}_{-}(0,u)=\frac{\sqrt{1-u} - 1}{\sqrt{1-u}} \,.
\label{eq:C}
\end{equation}
Finally, by plugging \eqref{eq:C} in (\ref{lambda_d_not_uni}) 
we obtain the unique generating function of the survival probability 
for a discrete-time random walk in continuous space with
exponential jump distribution, which is
\begin{equation}\label{lambda_d_uni}
\overline{\phi}_{+}(\xi,u) = 
\frac{\sqrt{1-u}-1}{1-u} \exp(-\sqrt{1-u} \, \xi) + \frac{1}{1-u} \,,
\end{equation}
that is consistent with formula (32) in reference 
\cite{majumdar_etal-jsp-2006}.
From formula \eqref{lambda_d_uni} 
we also have the boundary conditions: 
\begin{equation}\label{SA_bound}
\overline{\phi}_{+}(0,u)=\frac{1}{\sqrt{1-u}} \,,
\quad 
\overline{\phi}'_{+}(0,u)=\frac{1-\sqrt{1-u}}{\sqrt{1-u}} \,,
\end{equation}
where $\overline{\phi}_{+}(0,u)$ 
is in agreement with the Sparre Anderson theorem, and 
\begin{equation}\label{bound_infty}
\overline{\phi}_{\infty}(u)=\frac{1}{1-u}\,.
\end{equation}

We show now the effectiveness of the discussed approach
based on the Sturm--Liouville systems in the same special case. 
If we consider the special case $k(x)=\e^{-|x|}/2$,
see, e.g., \cite{gutkowiczkrusin_etal-jsp-1978,kou_etal-aap-2003,
majumdar_etal-jsp-2006,dahlenburg_etal-jpa-2022},
then from \eqref{kdev}
we have that $k'_z(z,\xi)=-\sgn(z-\xi)\, \e^{-|z-\xi|}/2$
and from \eqref{discontinuity} that 
$p(z)=1$, with $z\ne 0$.
Therefore, 
by solving \eqref{Lk} with respect to $q(z)$
in all points except $z=\xi$ 
it results $q(z)=-1$.
This means that the survival probability $\phi_{n+1}(z)$
from \eqref{Lphi} is the solution of the equation
\be
\frac{d^2\phi_{n+1}}{dz^2} - \phi_{n+1}(z)=
-\phi_n(z) \,, 
\label{eq:exp}
\ee
and the Sparre Andersen theorem \eqref{SA} is a required boundary condition
in analogy with the literature studies  
reported in Section \ref{sec:intro}.
We introduce the generating function  
$\overline{\phi}(z,u)$, see definition in \eqref{genfun},
then from \eqref{eq:exp} we have that it solves the equation
\be
\frac{d^2 \overline{\phi}(z,u)}{dz^2}
- \overline{\phi}(z,u) + 1 = -u \,\overline{\phi}(z,u) \,.
\label{SLS_exp4a}
\ee 
By applying the Laplace transform with respect to $z$,
where $\widetilde{\overline{\phi}}(s,u)$ stands for the Laplace transform 
of $\overline{\phi}(z,u)$ with $s \in \CC$, 
formula \eqref{SLS_exp4a} reads
\be
s^2 \,\widetilde{\overline{\phi}}(s,u) - s \overline{\phi}(0,u)
- \overline{\phi}'(0,u) - \widetilde{\overline{\phi}}(s,u)+\frac{1}{s}
= -u \, \widetilde{\overline{\phi}}(s,u) \,,
\label{SLS_exp4b}
\ee
such that
\be
\widetilde{\overline{\phi}}(s,u) =
\frac{s^2 \, \overline{\phi}(0,u) + s \, \overline{\phi}'(0,u) - 1}
{s (s^2 + u -1)} \,.
\label{SLS_exp4bb}
\ee
By taking boundary conditions \eqref{SA_bound},
we have
\begin{eqnarray}
\widetilde{\overline{\phi}}(s,u)
&=& \frac{s^2 + s(1-\sqrt{1-u)} - \sqrt{1-u}}
{s \, \sqrt{1-u} \, (s^2 +u -1)} \nonumber \\
&=& \frac{s(s+1) - (s+1)\sqrt{1-u)}}
{s \, \sqrt{1-u} \, (s^2 +u -1)} \nonumber \\
&=& \frac{s(s+1) - (s+1)\sqrt{1-u)}}
{s \, \sqrt{1-u} \, (s + \sqrt{1-u})(s-\sqrt{1-u})} \nonumber \\
&=& \frac{s+1}{s \,\sqrt{1-u} \, (s+\sqrt{1-u})} \,,
\label{SLS_exp4c}
\end{eqnarray}
from which we recover formula \eqref{lambda_d_uni} 
after the Laplace inversion.

\section{An example from the theory of the CTRW}
\label{sec:CTRW}
We discuss now the feasibility 
of the proposed approach based on the Sturm--Liouville system
through the same example but from the theory of the CTRW.
To this aim, we introduce the time variable $t \in \RR^+$ 
and random waiting-times between two consecutive jumps $\tau_j \in \RR_0^+$,
with $j \in \NN$, 
such that if the process starts at $t=0$ it holds 
$t=\sum_{j=1}^n \tau_j$.
The waiting-times $\tau_j$ are assumed to be
i.i.d. random variables distributed according to
a density function $\psi(t)$.
Moreover, we introduce also the probability that a 
given waiting interval between two consecutive jumps 
is greater than or equal to $t$, 
namely $\displaystyle{\Psi(t)=1-\int_0^t \psi(\tau) d\tau}$.
In this continuous-time setting,
we replace the notation of the survival probability 
in discrete epochs $\phi_n(\xi)$ with
the new notation $\Lambda(\xi,t)$ and now \eqref{WH} reads 
\cite[formula (16)]{dahlenburg_etal-jpa-2022}
\be
\lLambda(\xi,s)=\lPsi(s) + \lpsi(s)
\int_0^\infty k(z-\xi)\lLambda(z,s) \, dz \,,
\label{WHctrw}
\ee
where $\lLambda(\xi,s)$ stands for the Laplace transform 
of $\Lambda(\xi,t)$ with $s \in \CC$, 
and it holds $s\lPsi(s)=1-\lpsi(s)$. After Laplace inversion
and by setting $t=n \Delta t$ and $\psi(\tau)=\delta(\tau-\Delta t)$,
then $\Psi(0)=1$ and 
$\displaystyle{\Psi(t \ge \Delta t)=1-\int_0^t \delta(\tau-\Delta t)d\tau=0}$,
such that from \eqref{WHctrw} we recover \eqref{WH} as expected. 

Equation \eqref{WHctrw} is again a Fredholm integral equation
of the second kind with respect to $z$ and it can be solved
by methods developed for Wiener--Hopf integral equations 
\cite{estrada_kanwal-2000,kisil_etal-prsa-2021}.
Thus, by following Wiener--Hopf technique, 
we extend the interval of $\xi$ to $\RR$ and we
refer to $\lLambda_+(\xi,s)$ and $\lLambda_-(\xi,s)$ 
for {\it positive} and {\it negative} values of $\xi$, respectively,
such that $\lLambda=\lLambda_++\lLambda_-$.
We are interested to find $\lLambda_+(\xi,s)$ that corresponds
to the physical solution of \eqref{WHctrw} while $\lLambda_-(\xi,s)$ 
is an auxiliary function. Therefore, we study the following equation
\be
\lLambda(\xi,s)=\lPsi(s) + \lpsi(s)
\int_0^\infty k(z-\xi)\lLambda_+(z,s) \, dz \,,
\label{WHctrw2}
\ee
with $\lLambda_-(\xi,s)= \lpsi(s)
\int_0^\infty k(z-\xi)\lLambda_+(z,s) dz$.

We consider now, as a meaningful and manageable example, 
the exponential kernel $k(x)=\e^{-|x|}/2$, again,
see, e.g., \cite{gutkowiczkrusin_etal-jsp-1978,kou_etal-aap-2003,
majumdar_etal-jsp-2006,dahlenburg_etal-jpa-2022}.
By applying the Fourier transform \eqref{fourier} to \eqref{WHctrw2} we have
\be
\lfLambda_+(\omega,s)=
-
\frac{\omega(1-i\omega)\lLambda_-(0,s) + i(1+\omega^2)\lPsi(s)} 
{\omega(s\lPsi(s)+\omega^2)}
\,,
\label{lambdaeq0}
\ee
where by exploiting the exponential kernel we have 
$\fk(\omega)=1/(1+\omega^2)$ and from 
the definition of $\lLambda_-(\xi,s)$ 
it results $\lfLambda_-(\omega,s)=\lLambda_-(0,s)/(1+i\omega)$. 
We can invert \eqref{lambdaeq0} 
through \eqref{fourierinv} by using the residue theorem at the 
simple poles
$\omega=0$, $\omega=i\sqrt{s\lPsi(s)}$ and $\omega=-i\sqrt{s\lPsi(s)}$. 
Finally, we have
\begin{eqnarray}
\lLambda_+(\xi,s)=
\frac{1}{s}-
\left[\frac{\sqrt{s\lPsi(s)}(\sqrt{s\lPsi(s)}+1)}{2s\lPsi(s)}\lLambda_-(0,s)
+\frac{1-s\lPsi(s)}{2s}\right]\e^{\xi\sqrt{s\lPsi(s)}}
\nonumber \\
- \left[\frac{\sqrt{s\lPsi(s)}(\sqrt{s\lPsi(s)}-1)}{2s\lPsi(s)}\lLambda_-(0,s)
+\frac{1-s\lPsi(s)}{2s}\right]\e^{-\xi\sqrt{s\lPsi(s)}}
\,.
\label{lambdaeq1}
\end{eqnarray}
Hence, the survival probability $\lLambda_+(\xi,s)$ cannot be
determined form \eqref{lambdaeq1}, and consequently from
\eqref{WHctrw} and \eqref{WHctrw2}, because $\lLambda_-(0,s)$
is unknown. 

From the theory of CTRW we also have \cite{gorenflo_etal-csf-2007}
\be
\lLambda_+(\xi,s)=\lPsi(s) \sum_{n=0}^\infty
\phi_n(\xi) \, \left[\lpsi(s)\right]^n \,, 
\label{lambdactrw}
\ee
where now $n$ counts the number of occurred jumps.
If we set $\xi=0$ in \eqref{lambdactrw}, 
then from \eqref{SA} we can have $\phi_n(0)$.
Since \eqref{SA} holds for Markovian random walks,
in the framework of CTRW we have that the process is Markovian
when \cite{zwanzig-jsp-1983,mainardi_etal-pa-2000}
\be
\psi(\tau)=\psi_{\rm{M}}(\tau)=\e^{-\tau} \,,
\quad 
\lpsiM(s)=\lPsiM(s)=\frac{1}{1+s} \,,
\label{ass:M}
\ee
and it holds
\begin{eqnarray}
\lLambda_+(0,s)
&=& \lPsiM(s) \sum_{n=0}^\infty
\phi_n(0) 
\left[\lpsiM(s)\right]^n 
\nonumber \\
&=&\frac{\lPsiM(s)}{\sqrt{s\lPsiM(s)}}=\frac{1}{\sqrt{s(1+s)}} \,.
\label{SActrw}
\end{eqnarray}
In the long-time limit, i.e., $s \to 0$, we have
$\lLambda_+(0,s) \sim 1/\sqrt{s}$ that after Laplace inversion gives
$\Lambda_+(0,t) \sim  1/\sqrt{t}$ for $t \to +\infty$
\cite{artuso_etal-pre-2014}.
The above limit is the continuous-time counter-part of \eqref{SA}.
This limit can be indeed obtained from \eqref{SActrw}
with any distribution $\psi(\tau)$ such that 
$\lpsi(s) \sim 1 - s$ for $s \to 0$, that is a weak rearrangement of
Markovianity in the sense of the existence of a 
finite time-scale that gives a finite mean for the waiting-times.
Finally, by equalling \eqref{lambdaeq1} with $\xi=0$
and \eqref{SActrw} we can determine $\lLambda_-(0,s)$ as follows
\be
\lLambda_-(0,s)=\lPsiM(s) \, \frac{\sqrt{s\lPsiM(s)}-1}{\sqrt{s\lPsiM(s)}}
= \frac{\sqrt{s}-\sqrt{1+s}}{\sqrt{s}(1+s)} \,.
\label{lambdaO}
\ee
We can now complete the calculation of $\lLambda_+(\xi,s)$ in
the Markovian case that results to be 
\be
\lLambda_+(\xi,s)=
\frac{1}{s}\left\{
1-\left[1-\sqrt{s\lPsiM(s)}\right]\e^{-\xi\sqrt{s\lPsiM(s)}} \, \right\} \,,
\label{lambdafinal}
\ee
with
\be
\lLambda_+(0,s)=
\frac{\sqrt{1-\lpsiM(s)}}{s} \,, \quad
\lLambda'_+(0,s)=
\frac{\sqrt{1-\lpsiM(s)}-1+\lpsiM(s)}{s} \,,
\label{BC:CTRW}
\ee
and it is in agreement with a previous derivation on the basis of
probabilistic arguments \cite{kou_etal-aap-2003}.
In fact, by introducing the quantity 
$\llambda(\xi,s)=1- s \, \lLambda_+(\xi,s)$
we have
\be
\llambda(\xi,s)=
\left[1-\sqrt{s\lPsiM(s)}\right]\e^{-\xi\sqrt{s\lPsiM(s)}} \,,
\ee
that is the formula obtained by S.G. Kou and H. Wang 
for the same simple case considered here
\cite[Theorem 3.1, formula (3.1)]{kou_etal-aap-2003}.

In analogy with the previous section, we show that the proposed
Sturm--Liouville system can be used for calculating the survival probability also
for CTRW models. In this case, instead to consider the generating
function, we consider formula \eqref{lambdactrw} that allows for
stepping from the discrete-time to the continuous-time setting.
By taking into account \eqref{Lphi}, we have in the most general case
that $\widetilde{\Lambda}_+(z,s)$ satisfies the equation
\be
\frac{d}{dz}\lkk p(z) \, \frac{d}{dz} \widetilde{\Lambda}_+(z,s) \rkk 
+ q(z) \, \widetilde{\Lambda}_+(z,s) 
= - \widetilde{\psi}(s) \, \widetilde{\Lambda}_+(z,s) 
+ q(z) \, \widetilde{\Psi}(s)\,.
\label{SLS_CTRW2}
\ee
Hence, in the special case $k(x)=\e^{-|x|}/2$,
see, e.g., \cite{gutkowiczkrusin_etal-jsp-1978,kou_etal-aap-2003,
majumdar_etal-jsp-2006,dahlenburg_etal-jpa-2022}, 
it holds $p(z)=1$ and $q(z)=-1$ and,
together with the Markovian assumption \eqref{ass:M}, 
equation \eqref{SLS_CTRW2} reduces to
\be
\frac{d^2 \widetilde{\Lambda}_+}{dz^2} 
- \widetilde{\Lambda}_+(z,s) 
= - \lpsiM \, \widetilde{\Lambda}_+(z,s) 
- \lPsiM(s)\,.
\label{SLS_CTRW3}
\ee
To conclude, by applying the Laplace transform to $z$ into $v$
it holds
\be
\widetilde{\widetilde{\Lambda}_+}(v,s) 
= \frac{v \,\widetilde{\Lambda}_+'(0,s) 
+ v^2 \, \widetilde{\Lambda}_+(0,s)
-\lPsiM(s)}{v \,\lk v^2-1+\lpsiM(s)\rk}\,,
\ee
and, after plugging boundary conditions \eqref{BC:CTRW},
we have 
\be
\widetilde{\widetilde{\Lambda}_+}(v,s) 
= \frac{(v+1) \, \sqrt{1-\lpsiM(s)}}{vs \,\lk v + \sqrt{1-\lpsiM(s)}\rk}
\,.
\label{SLS_CTRW4} 
\ee
Finally, it is possible to check that, by the Laplace inversion in $v$,
from \eqref{SLS_CTRW4} we can recover \eqref{lambdafinal}.

\section{Conclusion}
\label{sec:conclusions}
In this paper we derived the Sturm--Liouville system of equations
that is associated with the Wiener--Hopf integral equation \eqref{WH}
for the calculation of the survival probability in first-passage time
problems for random walks.
By studying a simple, but meaningful, example we showed the feasibility
of the proposed approach within the framework of 
the discrete-time random walk and that of the CTRW.
This approach is an alternative to the 
Pollaczek--Spitzer formula.

Summarising, we showed that $k(z,\xi)$ and $\phi_{n+1}(z)$ 
are indeed related accordingly to \eqref{WH}
if and only if they solve \eqref{Lk} and \eqref{Lphi}, respectively,
and boundary conditions \eqref{BC} are met.
Hence, from the derived Sturm--Liouville system \eqref{SL}
the Sparre Andersen theorem \eqref{SA} cannot be obtained 
but it is indeed a necessary boundary condition.
In other words, 
from the simple and concrete example here studied,
it follows that when the survival probability is calculated
by 
the proposed Sturm--Liouville system \eqref{SL}
an extra condition must be added, and 
this extra condition is actually the statement of 
the Sparre Andersen theorem \eqref{SA}.
As we discussed in the Introduction, 
this is a limitation also in the use of the Pollaczek--Spitzer formula
in its original derivation \cite{spitzer-1957} and analogue
to the assumption of the boundedness of the survival probability
in a more cumbersome derivation 
\cite{ivanov-aa-1994,comtet_etal-jsm-2005,majumdar_etal-jsm-2014}.

We end by underlying that the proposed approach 
allows for deriving explicit formulae for the survival probability 
in the discrete-time setting and also in the continuous-time one,
this last feasibility supports to extend its successful application 
even to non-Markovian CTRW models.
\vskip6pt


\enlargethispage{20pt}

\ack{This research is supported by the Basque Government through 
the BERC 2022--2025 program, 
by the Ministry of Science and Innovation: BCAM Severo Ochoa accreditation CEX2021-001142-S / MICIN / AEI / 10.13039/501100011033,
and through the Predoc Severo Ochoa 2018 grant PRE2018-084427.
}


\vskip2pc

\bibliographystyle{prsb} 
\bibliography{all} 

\end{document}